\documentclass[lettersize,journal]{IEEEtran}
\usepackage{amsmath,amsfonts}
\usepackage{algorithmic}
\usepackage{algorithm}
\usepackage{array}
\usepackage[caption=false,font=normalsize,labelfont=sf,textfont=sf]{subfig}
\usepackage{textcomp}
\usepackage{stfloats}
\usepackage{url}
\usepackage{verbatim}
\usepackage{graphicx}
\usepackage{cite}
\usepackage[hidelinks]{hyperref}

\usepackage{multirow} 
\usepackage{braket}
\usepackage{ragged2e}
\newtheorem{definition}{Definition}

\begin{document}

\title{Quantum Complex-Valued Self-Attention Model}

\author{Fu Chen, Qinglin Zhao, Li Feng, Longfei Tang, Yangbin Lin, Haitao Huang.
\thanks{Corresponding author: Qinglin Zhao.(e-mail: qlzhao@must.edu.mo)}
\IEEEcompsocitemizethanks{\IEEEcompsocthanksitem Fu Chen, Qinglin Zhao, Li Feng and Haitao Huang are with Faculty of Innovation Engineering, Macau University of Science and Technology, 999078, China. 
\IEEEcompsocthanksitem Longfei Tang is with College of Electrical Engineering and Automation, Fuzhou University. Fuzhou 350108, China.
\IEEEcompsocthanksitem Yangbin Lin is with College of Computer Engineering College, Jimei Univeristy, Xiamen 361021, China. 
}
}

\markboth{Journal of \LaTeX\ Class Files,~Vol.~14, No.~8, August~2021}%
{Shell \MakeLowercase{\textit{et al.}}: A Sample Article Using IEEEtran.cls for IEEE Journals}


\maketitle

\begin{abstract}
  Self-attention has revolutionized classical machine learning, yet existing quantum self-attention models underutilize quantum states' potential due to oversimplified or incomplete mechanisms. To address this limitation, we introduce the Quantum Complex-Valued Self-Attention Model (QCSAM), the first framework to leverage complex-valued similarities, which captures amplitude and phase relationships between quantum states more comprehensively. To achieve this, QCSAM extends the Linear Combination of Unitaries (LCUs) into the Complex LCUs (CLCUs) framework, enabling precise complex-valued weighting of quantum states and supporting quantum multi-head attention. Experiments on MNIST and Fashion-MNIST show that QCSAM outperforms recent quantum self-attention models, including QKSAN, QSAN, and GQHAN. With only 4 qubits, QCSAM achieves 100\% and 99.2\% test accuracies on MNIST and Fashion-MNIST, respectively. Furthermore, we evaluate scalability across 3-8 qubits and 2-4 class tasks, while ablation studies validate the advantages of complex-valued attention weights over real-valued alternatives. This work advances quantum machine learning by enhancing the expressiveness and precision of quantum self-attention in a way that aligns with the inherent complexity of quantum mechanics.

\end{abstract}


\section{Introduction}
The self-attention mechanism, as a key component of deep learning architectures, has significantly impacted the ways in which data is processed and features are learned \cite{vaswani2017attention,zong2024self,xu2023multimodal}. By generating adaptive attention weights, self-attention not only highlights key features in the data but also integrates global contextual information, thereby improving the expressive power and computational efficiency of deep learning systems. For instance, in natural language processing \cite{brown2020language,devlin2018bert,radford2018improving}, self-attention has enhanced language understanding and generation by capturing long-range dependencies and contextual information; in computer vision \cite{han2022survey, li2022contextual,chen2024survey}, it allows models to focus on key regions within images to optimize feature extraction; and in recommender systems \cite{zhang2024ninerec,zhang2023personalized}, it improves the accuracy of capturing user behavior and preferences, thereby enhancing the effectiveness of personalized recommendations. Large-scale models such as GPT-4 \cite{achiam2023gpt} have further exploited the potential of self-attention, allowing them to address multimodal tasks such as visual question answering, image captioning, and cross-modal reasoning. These developments demonstrate that the self-attention mechanism is a fundamental mechanism of deep learning’s success and motivates the exploration of similar mechanisms in quantum machine learning.

Inspired by the success of self-attention mechanisms in classical deep learning, and with the rapid progress in quantum computing \cite{preskill2018quantum,zhou2020limits}, quantum self-attention models have emerged as a quantum adaptation of classical attention mechanisms. These models seek to investigate the application of quantum systems’ unique properties within self-attention frameworks, facilitating new research areas in quantum machine learning \cite{biamonte2017quantum,schuld2019quantum,shi2022parameterized,dunjko2016quantum}. This development has attracted significant attention by combining the representational power of self-attention with the computational benefits of quantum technologies.

\subsection{Motivation}
Quantum attention weights are a fundamental component of quantum self-attention models, where effectively utilizing quantum computational advantages is essential for their performance. Currently, there are several approaches for calculating these weights, but each has limitations. One approach involves fusion-based methods \cite{zhao2024qksan}, which attempt to combine  the query state $\ket{Q}$ and the key state $\ket{K}$ to estimate their similarity. However, these methods often rely on simplified fusion processes, such as simple logical gates like CNOT or parameterized circuits, which may not fully capture the complex interactions between quantum states.
Another approach employs real-valued overlap methods \cite{li2024quantum,chen2025quantum}, which transform similarity computations into real-valued overlaps. However, this transformation does not preserve the phase information that is fundamental to quantum states. Quantum states are inherently complex-valued, and their phase differences drive quantum interference, which is central to the computational power of quantum systems. By neglecting the phase, these models limit their expressive capacity. Furthermore, implicit relationship \cite{zhao2024gqhan} methods avoid explicit pairwise similarity computations between $\ket{Q}$ and $\ket{K}$. Instead, they employ a trainable circuit to directly compute target weights, thereby bypassing the extraction of explicit pairwise interaction details. This design may reduce interpretability and fail to capture the pairwise information in quantum state interactions.

To leverage the advantages of quantum computing, we aim to develop a quantum self-attention mechanism that utilizes the complex inner product between quantum states to measure their similarity. This inner product inherently captures the similarity in both the real and imaginary parts, which indirectly reflects their magnitude and phase relationships. By doing so, our approach enables the creation of more precise and expressive quantum self-attention models that fully exploit the quantum nature of the data.

\subsection{Contributions}
Current quantum self-attention models, when leveraging the expressive power of quantum states, are often limited by their dependence on real-valued overlaps or simplistic fusion methods for attention weights, which fail to fully utilize the complex-valued nature of quantum states. To address this limitation, we introduce the Quantum Complex-Valued Self-Attention Model (QCSAM), which employs a complex-valued attention mechanism to comprehensively capture the relationships between quantum states. This innovation significantly enhances the precision and expressive power of quantum self-attention models, offering a novel approach in quantum machine learning that aligns with the intrinsic principles of quantum mechanics. The main contributions of this work are as follows:

\begin{itemize}
    \item We introduce the Quantum Complex-Valued Self-Attention Model (QCSAM), the first framework to derive complex-valued attention weights from the real and imaginary parts of $\braket{K|Q}$. This approach captures the amplitude and phase relationships between quantum states, enabling a precise representation of quantum similarity consistent with the complex nature of quantum mechanics.

    \item We enhance our quantum self-attention model by generalizing Linear Combination of Unitaries (LCUs) to Complex Linear Combination of Unitaries (CLCUs), enabling the incorporation of complex coefficients. This generalization assumes quantum self-attention weights are complex, introducing a prior preference that aligns with the complex-valued nature of quantum systems. Leveraging the CLCUs framework, we introduce a quantum multi-head self-attention mechanism, where each head independently learns complex weights, further strengthening the model’s representational capacity.

    \item We conducted thorough evaluations of the proposed Quantum Complex Self-Attention Model (QCSAM) on the MNIST and Fashion-MNIST datasets, demonstrating its superior classification accuracy compared to existing quantum self-attention models (e.g., QKSAN, QSAN, GQHAN). On a 4-qubit system, QCSAM achieved test accuracies of 100\% and 99.2\% for MNIST and Fashion-MNIST, respectively. Scalability studies investigated the effects of varying qubit counts and task complexity on performance, offering insights into the model’s behavior across different configurations. Notably, the dual-head attention configuration consistently outperformed the single-head attention configuration across all evaluated tasks, underscoring its advantage in improving classification performance. Additionally, ablation studies confirmed that employing complex-valued attention weights significantly enhances performance compared to using real-valued attention weights.
\end{itemize}

\section{Preliminaries and Related Work}
\begin{figure*}[!t]
  \centering
  \includegraphics[width=7in]{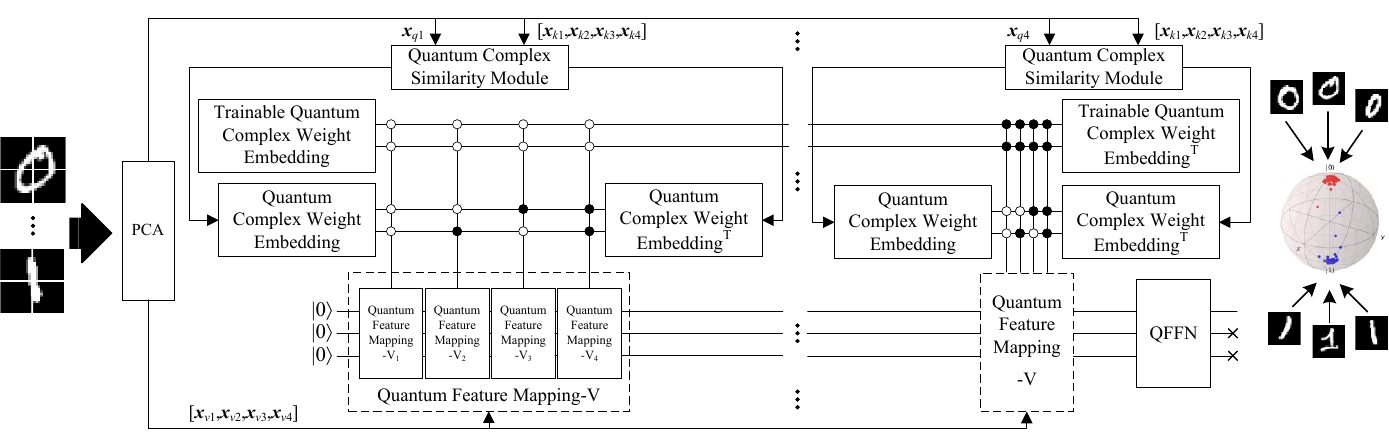}
  \caption{The framework of the proposed Quantum Complex-Valued Self-Attention Model (QCSAM).}
  \label{framework}
\end{figure*}

This section introduces the basic concepts of quantum machine learning involved in the paper, laying the foundation for the subsequent theoretical derivation and model construction.

\subsection{Preliminaries}
\subsubsection{Pure Quantum State}
A pure quantum state represents a system that is fully described by a single state vector, without uncertainty in its properties. It is represented by a state vector $\ket{\psi}$ in a Hilbert space and satisfies the normalization condition $\braket{\psi|\psi}=1$. Under a quantum operation, the state evolves as $\ket{\psi'}=U\ket{\psi}$, where $U$ is a unitary operator that meets the requirement $U^\dagger U=I$.

\subsubsection{Mixed Quantum State}
A mixed quantum state describes a system that is in a probabilistic mixture of different quantum states, rather than being in a single pure state. It is represented by a density matrix $\rho$, which is semi-positive definite, Hermitian, and satisfies $\text{Tr}(\rho) = 1$. 
The density matrix is defined as $\rho = \sum_j p_j \ket{\psi_j}\bra{\psi_j}$, representing an ensemble of pure states $\{p_j, \ket{\psi_j}\}$ with probabilities $p_j$. Under a quantum operation, the state evolves according to $\rho' = U \rho U^\dagger$.

\subsubsection{Standard LCUs Method}
The Linear Combination of Unitaries (LCUs) method implements a weighted sum of multiple unitary operations by preparing an ancilla superposition and executing controlled operations \cite{childs2017quantum, kothari2014efficient, berry2015simulating}. Specifically, the preparation operation $U_{\text{PREP}}$ transforms the ancilla register from its initial state $\ket{b_1}$ into the superposition state:
\begin{equation} 
    U_{\text{PREP}} \ket{0}^{\otimes n} = \frac{1}{\sqrt{\mathcal{N}}} \sum_{j=0}^{N-1} \sqrt{\alpha_j} \ket{j},
\end{equation}
where for simplicity $\alpha_j \geq 0$, $\alpha_j \in \mathbb{R}$, $\mathcal{N}$ is the normalization constant, and $\ket{j}$ denotes the computational basis states. Next, the selection operation $U_{\text{SELECT}}$ conditionally applies the corresponding unitary $U_j$ to the target state $\ket{\psi}$ based on the ancilla state:
\begin{equation} 
U_{\text{SELECT}} \ket{j}\ket{\psi} = \ket{j} U_j \ket{\psi}.
\end{equation}

Subsequently, the inverse preparation operation \( U_{\text{PREP}}^\dagger \) is applied, yielding:
\begin{equation}
\begin{aligned}
    &(U_{\text{PREP}}^\dagger \otimes I_{\text{target}}) U_{\text{SELECT}} (U_{\text{PREP}} \otimes I_{\text{target}}) \ket{0}^{\otimes n} \ket{\psi} \\
    = &\frac{1}{\mathcal{N}} \ket{0}^{\otimes n}  \sum_{j=0}^{N-1} \alpha_j U_j \ket{\psi} + \text{orthogonal terms},
\end{aligned}
\end{equation}
where $I_{\text{target}}$ is the identity operator on the Hilbert space $\mathcal{H}_{\text{target}}$ in which $|\psi\rangle$ resides. The “orthogonal terms” correspond to the components in the ancilla space that are orthogonal to $\ket{0}^{\otimes n}$. Finally, if the ancilla register is measured and the outcome is $\ket{0}^{\otimes n}$, the target state is projected to:
\begin{equation} 
  \begin{aligned}
    &\bra{0}^{\otimes n}(U_{\text{PREP}}^\dagger \otimes I_{\text{target}}) U_{\text{SELECT}} (U_{\text{PREP}} \otimes I_{\text{target}}) \ket{0}^{\otimes n} \ket{\psi} \\
    =&\frac{1}{\mathcal{N}'} \sum_{j=0}^{N-1} \alpha_j U_j \ket{\psi},
  \end{aligned}
\end{equation}
where $\mathcal{N}' = \sqrt{P_{\text{success}}}\mathcal{N}$ is the normalization constant and measuring the outcome $\ket{0}^{\otimes n}$ occurs with success probability $P_{\text{success}} = \frac{1}{\mathcal{N}^2} | \sum_{j=0}^{N-1} \alpha_j U_j \ket{\psi} |^2$. This process effectively implements the operation:
\begin{equation} 
    A = \frac{1}{\mathcal{N}'} \sum_{j=0}^{N-1} \alpha_j U_j.
\end{equation}

\subsection{Related Work}
Existing quantum self-attention models are generally classified into two categories based on the source of their trainable parameters. One category consists of hybrid models, in which trainable parameters are derived from both quantum circuits and classical modules. In contrast, the other category comprises models in which all trainable parameters are derived exclusively from quantum circuits.

For models integrating quantum and classical computing for trainable parameters, the key idea is to leverage the expressive power of quantum states alongside the flexibility of classical computing. For example, the Quantum Self-Attention Neural Network (QSANN) \cite{li2024quantum} employs parameterized quantum circuits (PQC) \cite{haug2021capacity,sim2019expressibility,benedetti2019parameterized} to generate the Query-Key-Value (Q-K-V) representations in a classical self-attention mechanism, while self-attention weights are computed using a classical Gaussian function. The Quixer framework \cite{khatri2024quixer} processes inputs and outputs through classical neural network but incorporates quantum modules such as linear combination unitaries (LCUs) \cite{chakraborty2024implementing,berry2015hamiltonian} and quantum singular value transformation (QSVT) \cite{gilyen2019quantum,low2019hamiltonian} to construct a quantum-enhanced Transformer. Furthermore, the Quantum Mixed-State Self-Attention Network (QMSAN) \cite{chen2025quantum} encodes classical inputs into mixed quantum states using a trainable quantum embedding circuit. Then it derives attention weights by applying the SWAP test \cite{garcia2013swap,zhao2019building,fanizza2020beyond} to mixed quantum states $\rho_Q$ and $\rho_K$ and classically combines these weights with the measurement results of the quantum state $\ket{V}$ to finalize the self-attention mechanism.

The second category of quantum self-attention models is characterized by the implementation of all trainable parameters entirely within quantum circuits, leveraging the inherent advantages of quantum computing to achieve a fully quantum algorithmic realization. For example, the Quantum Self-Attention Network (QKSAN) \cite{zhao2024qksan} model uses quantum kernel methods \cite{paine2023quantum} to compute attention weights between quantum states $\ket{Q}$ and $\ket{K}$, followed by the integration of the quantum value state $\ket{V}$ through gate operations $C(Ry)$. Another notable approach in \cite{guo2024quantum} proposes a method for adapting Transformer models to quantum settings by embedding pretrained classical parameters into quantum circuits. This approach enables efficient implementation of the core Transformer module using quantum matrix operations, allowing inference to be performed on quantum computers. The Grover-inspired Quantum Hard Attention Network (GQHAN) \cite{zhao2024gqhan} model introduces quantum hard attention based on the Grover algorithm \cite{long2001grover,godfrin2017operating}, bypassing the calculation of the similarity of Q and K in traditional self-attention mechanisms. Instead, it leverages an oracle and diffusion operator to amplify key information within quantum states. Additionally, Quantum Vision Transformers \cite{cherrat2024quantum} exploit the properties of orthogonal quantum layers to efficiently execute the linear algebra operations necessary for quantum computing.

A common characteristic of these quantum self-attention models is their oversimplified or incomplete calculations of similarity between quantum states. Our work design explicitly leverages the complex nature of quantum states in the self-attention mechanism, incorporating both amplitude and phase information for a more comprehensive representation of quantum states relationships.

\section{Methodology}
In this section, we begin by introducing the framework and providing the theoretical motivation for adopting Complex Linear Combination of Unitaries (CLCUs). Following this, we describe the design and functionality of the core modules and the loss function.
\subsection{General Framework}
Figure \ref{framework} illustrates the architecture of QCSAM. The process begins by dividing the input data into smaller patches, each of which is subsequently reduced in dimensionality using Principal Component Analysis (PCA) \cite{abdi2010principal}. These reduced classical data patches are processed through two parallel pathways. In the first pathway, the Quantum Feature Mapping (QFM) module transforms each reduced patch into a quantum state $\ket{V_j}$. Simultaneously, in the second pathway, the Quantum Complex Similarity Module (QCS) employs two QFM submodules to transform each patch into quantum states $\ket{Q_k}$ and $\ket{K_j}$. The QCS module computes the inner products $\braket{K_j|Q_k}$ as complex self-attention weights between each query state $\ket{Q_k}$ and all key states $\ket{K_j}$. These weights are applied by the Quantum Complex Weighting (QCWE) module to aggregate the corresponding value states $\ket{V_j}$ using a Complex Linear Combination of Unitaries (CLCU) framework, yielding a weighted sum for each query. Additionally, the CLCUs framework incorporates a Trainable QCWE component, which introduces learnable complex weights optimized during training to further refine the aggregation of quantum value states. The resulting output is processed through a Quantum Feedforward Network (QFFN), which integrates global context into the feature representation. Finally, the resulting quantum states are measured to produce classical outputs for the classification task. In the classical self-attention mechanism, the attention is computed as $\text{Attention}(Q, K, V) = \text{softmax}\left(\frac{QK^T}{\sqrt{d_k}}\right) V$. We propose a quantum version of self-attention, defined as:
\begin{equation}
\begin{aligned}
    &\text{QAttention}\left( \{\ket{Q_k}\}_{k=0}^{N-1}, \{\ket{K_j}\}_{j=0}^{N-1}, \{\ket{V_j}\}_{j=0}^{N-1} \right) \\
    =& \left\{ \frac{1}{\mathcal{N}_{S_{k}}} \sum_{j=0}^{N-1} \braket{K_j | Q_k} \ket{V_j} \right\}_{k=0}^{N-1},
\end{aligned}
\end{equation}
where $\braket{K_j|Q_k}$ represents complex numbers. 

In the context of implementing our quantum attention mechanism, the standard Linear Combination of Unitaries (LCUs) approach presents limitations when handling the inherently complex-valued nature of quantum computations. The LCUs method requires real-valued coefficients $\alpha_j$, which implicitly absorb phase information into modified unitary operators $U'_j$ (via $\alpha_jU_j = |\alpha_j|e^{i\phi_j}U_j = |\alpha_j|U'_j$). While mathematically equivalent, this real-valued parameterization introduces a practical limitation because requiring specially designed unitary operators $U_j$ to account for phase effects. This design leads to increased circuit complexity and constrained flexibility in capturing amplitude-phase relationships.

In contrast, our Complex Linear Combination of Unitaries (CLCUs) framework leverages complex coefficients $\alpha_j \in \mathbb{C}$ to weight arbitrary unitary operators $U_j$. This approach explicitly embeds both amplitude and phase within the coefficients themselves, eliminating the need for additional phase adjustments in $U_j$ and simplifying quantum circuit design. By assuming complex-valued quantum self-attention weights, CLCUs enable optimization in the complex domain. This introduces an inductive bias that reflects quantum mechanics’ reliance on complex Hilbert spaces. As a result, this approach reduces circuit complexity, and increases representational capacity by effectively capturing the intricate amplitude-phase relationships inherent to quantum systems.

\subsection{Quantum Feature Mapping Module}
Quantum Feature Mapping is the first step in the quantum machine learning pipeline, where classical input data is transformed into quantum states. The design of this mapping process is critical, as it directly affects the performance and representational capacity of subsequent quantum circuits. We employ a trainable quantum embedding architecture \cite{lloyd2020quantum}. This design introduces flexibility in quantum feature representation by incorporating trainable parameters, thereby improving model performance. The architecture is defined as follows:
\begin{equation}
    U_{\text{QFM}}(\boldsymbol{x},\boldsymbol{\theta})=V(\boldsymbol{x})\prod_{l=1}^{L} (W_l( \boldsymbol{\theta}_l)V(\boldsymbol{x})),
\end{equation}
where $V(\boldsymbol{x})$ represents the data encoding layer, and $W_l(\boldsymbol{\theta}_l)$ denotes the variational layer at depth $l$, with $\boldsymbol{\theta}_l$ as the trainable parameters.
Starting from an initial state $\ket{0}^{\otimes n}$, the circuit maps the classical input feature vector $\boldsymbol{x} = (x_1, x_2, \dots, x_N)^T$ into the quantum domain using $V(\boldsymbol{x})$, implemented via single-qubit $R_x$ rotation gates with angles proportional to $x_i$. Each $W_l(\boldsymbol{\theta})$ consists of parameterized two-qubit $ZZ$ gates, which control the entanglement between qubits, and single-qubit $R_y$ rotations, enhancing the circuit’s expressive power. This structure alternates between $V(x)$ and $W_l(\boldsymbol{\theta}_l)$ on the $L$ layers, concluding with a final $V(\boldsymbol{x})$ layer, enabling the construction of rich quantum representations through repeated data encoding and entanglement.

For the self-attention mechanism, distinct states are generated using separate parameter sets:
\begin{equation}
\begin{split}
        \ket{Q}=U_{\text{QFM}}(\boldsymbol{x},\boldsymbol{\theta_Q})\ket{0}^{\otimes n}, \\\ket{K}=U_{\text{QFM}}(\boldsymbol{x},\boldsymbol{\theta_K})\ket{0}^{\otimes n}, \\\ket{V}=U_{\text{QFM}}(\boldsymbol{x},\boldsymbol{\theta_V})\ket{0}^{\otimes n}.
\end{split}
\end{equation}

For details on the architecture of the Quantum Feature Mapping Module, please refer to the supplementary file \ref{section: Circuit Architecture}.

\subsection{Quantum Complex Similarity Module}
Building upon the quantum feature mapping, the similarity between states $\ket{Q}$ and $\ket{K}$ defines the quantum self-attention weights. Traditional methods, such as the SWAP test and quantum kernel approaches, compute $|\braket{Q|K}|^2$, capturing magnitude but neglecting the phase, which is essential for encoding quantum state relationships. To address this, we propose an enhanced Hadamard test circuit that measures both real and imaginary parts of $\braket{Q|K}$ using selection, auxiliary, and working qubits. This method provides a complete similarity measure, incorporating magnitude and phase, thereby enhancing the model's ability to capture intricate quantum interactions.

\begin{definition}[Quantum Complex Self-Attention Weight]\label{def:inner_product} For two $n$-qubit states, $\ket{Q}$ and $\ket{K}$, we define their quantum complex self-attention weight as the inner product: 
\begin{equation} 
\begin{aligned} 
\braket{K|Q} &= \Biggl( \sum_{k=0}^{N-1} (c_k - d_k \mathrm{i}) \bra{k} \Biggr) \Biggl( \sum_{j=0}^{N-1} (a_j + b_j \mathrm{i}) \ket{j} \Biggr) \\ 
&= \sum_{k=0}^{N-1} \Bigl[(a_k c_k + b_k d_k) + \mathrm{i}(b_k c_k - a_k d_k)\Bigr]. \\
&= \text{Re}(\braket{K|Q}) + \mathrm{i}\text{Im}(\braket{K|Q}), 
\end{aligned} 
\end{equation} 
where $\mathrm{i}$ represents the imaginary unit. ${\ket{k}}$ and ${\ket{l}}$ represent the computational basis states for the $n$ qubits, with $\braket{k|l}=\delta_{kl}$, and the normalization conditions $\sum_{j=0}^{N-1} (a_j^2 + b_j^2)=1$, $\sum_{k=0}^{N-1} (c_k^2 + d_k^2)=1$. 
\end{definition}

\begin{figure}[h]
  \centering
  \includegraphics[width=3.5in]{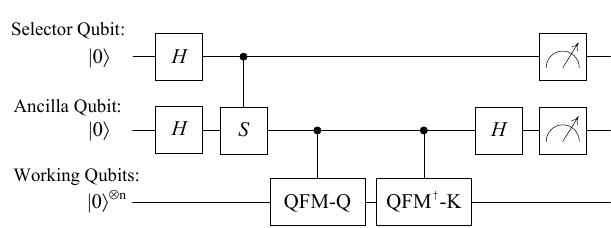}
  \caption{Enhanced Hadamard Test Circuit for Measuring Real and Imaginary Parts of Quantum Complex Self-Attention Weights.}
  \label{fig: Complex Hadamard Test Circuit}
  \end{figure}

To extract the real and imaginary parts of the quantum attention weights, we design an improved Hadamard test circuit with three types of qubits, as shown in fig. \ref{fig: Complex Hadamard Test Circuit}: 
\begin{itemize} 
\item \textbf{Selection Qubit ($q_0$):} Determines whether to measure the real or imaginary part. A measurement result of $\ket{0}$ means the real part will be measured, and $\ket{1}$ means the imaginary part will be measured. 
\item \textbf{Auxiliary Qubit ($q_1$):} Stores the measurement result corresponding to the selected component. When $q_0$ is $\ket{0}$, the expectation value measured on $q_1$ corresponds to $\text{Re}(\braket{K|Q})$; when $q_0$ is $\ket{1}$, it corresponds to $\text{Im}(\braket{K|Q})$. 
\item \textbf{Working Qubits ($q_2$):} Encodes the quantum states $\ket{Q}$ and $\ket{K}$ and operates on them. 
\end{itemize}

Assume the initial quantum state is: 
\begin{equation} 
\ket{\psi_0} = \ket{0}_0 \otimes \ket{0}_1 \otimes \ket{0}^{\otimes n}_2. 
\end{equation} 

After processing through the improved Hadamard test circuit, the state evolves as follows: 
\begin{align*} 
\ket{\psi_1} &= \frac{1}{\sqrt{2}} \ket{0}_0 \otimes \frac{1}{2}\Bigl[(\ket{0}+\ket{1})_1 \otimes \ket{0}^{\otimes n}_2 \\ 
&\quad + (\ket{0}-\ket{1})_1 \otimes U_{\text{QFM}}^\dagger(\boldsymbol{x},\boldsymbol{\theta}_K) U_{\text{QFM}}(\boldsymbol{x},\boldsymbol{\theta}_Q) \ket{0}^{\otimes n}_2\Bigr] \\ 
&\quad + \frac{1}{\sqrt{2}} \ket{1}_0 \otimes \frac{1}{2}\Bigl[(\ket{0}+\ket{1})_1 \otimes \ket{0}^{\otimes n}_2 \\ 
&\quad + \mathrm{i}(\ket{0}-\ket{1})_1 \otimes U_{\text{QFM}}^\dagger(\boldsymbol{x},\boldsymbol{\theta}_K) U_{\text{QFM}}(\boldsymbol{x},\boldsymbol{\theta}_Q) \ket{0}^{\otimes n}_2\Bigr], 
\end{align*} 
where $U_{\text{QFM}}(\boldsymbol{x},\boldsymbol{\theta}_Q)\ket{0}^{\otimes n}=\ket{Q}$ and $U_{\text{QFM}}(\boldsymbol{x},\boldsymbol{\theta}_K)\ket{0}^{\otimes n}=\ket{K}$, with the circuits $U_{\text{QFM}}(\boldsymbol{x},\boldsymbol{\theta}_Q)$ and $U_{\text{QFM}}(\boldsymbol{x},\boldsymbol{\theta}_K)$ generated by the Quantum Feature Mapping Module QFM-Q and QFM-K, respectively.

The measurement process proceeds as follows: 
\begin{itemize} 

\item \textbf{Selection of Real Part Component:} Measurement of $q_0$ yields $\ket{0}$. 

The state collapses to: 
\begin{equation} 
\begin{split} 
\ket{\psi_2} = &\frac{1}{\sqrt{2}} \Bigl[ \ket{0}_1 \otimes \Bigl(\ket{0}^{\otimes n}_2 \\
&+ U_{\text{QFM}}^\dagger(\boldsymbol{x},\boldsymbol{\theta}_K) U_{\text{QFM}}(\boldsymbol{x},\boldsymbol{\theta}_Q) \ket{0}^{\otimes n}_2\Bigr) \\
&+ \ket{1}_1 \otimes \Bigl(\ket{0}^{\otimes n}_2 \\
&- U_{\text{QFM}}^\dagger(\boldsymbol{x},\boldsymbol{\theta}_K) U_{\text{QFM}}(\boldsymbol{x},\boldsymbol{\theta}_Q) \ket{0}^{\otimes n}_2\Bigr) \Bigr]
\end{split} 
\end{equation}

Measuring the expectation value on $q_1$ gives: 
\begin{equation} 
\begin{split} 
P_0 &= \frac{1}{4}\left|\ket{0}^{\otimes n}_2 + U_{\text{QFM}}^\dagger(\boldsymbol{x},\boldsymbol{\theta}_K) U_{\text{QFM}}(\boldsymbol{x},\boldsymbol{\theta}_Q) \ket{0}^{\otimes n}_2\right|^2 \\ 
&= \frac{1 + \text{Re}(\bra{0}^{\otimes n}_2 U_{\text{QFM}}^\dagger(\boldsymbol{x},\boldsymbol{\theta}_K) U_{\text{QFM}}(\boldsymbol{x},\boldsymbol{\theta}_Q) \ket{0}^{\otimes n}_2)}{2} \\
&= \frac{1 + \text{Re}(\braket{K|Q})}{2}. 
\end{split} 
\end{equation}

\item \textbf{Selection of Imaginary Part Component:} Measurement of $q_0$ yields $\ket{1}$. 

The state becomes: 
\begin{equation} 
\begin{split} 
\ket{\psi_{3}} = &\frac{1}{\sqrt{2}} \Bigl[ \ket{0}_1 \otimes \Bigl(\ket{0}^{\otimes n}_2 \\
&+  \mathrm{i} \, U_{\text{QFM}}^\dagger(\boldsymbol{x},\boldsymbol{\theta}_K) U_{\text{QFM}}(\boldsymbol{x},\boldsymbol{\theta}_Q) \ket{0}^{\otimes n}_2\Bigr) \\
&+ \ket{1}_1 \otimes \Bigl(\ket{0}^{\otimes n}_2 \\
&-  \mathrm{i} \, U_{\text{QFM}}^\dagger(\boldsymbol{x},\boldsymbol{\theta}_K) U_{\text{QFM}}(\boldsymbol{x},\boldsymbol{\theta}_Q) \ket{0}^{\otimes n}_2\Bigr) \Bigr]
\end{split} 
\end{equation}

The measurement on $q_1$ gives: 
\begin{align*} 
P_0 &= \frac{1}{4}\left|\ket{0}^{\otimes n}_2 +  \mathrm{i} U_{\text{QFM}}^\dagger(\boldsymbol{x},\boldsymbol{\theta}_K) U_{\text{QFM}}(\boldsymbol{x},\boldsymbol{\theta}_Q) \ket{0}^{\otimes n}_2\right|^2 \\ 
&= \frac{1 - \text{Im}(\bra{0}^{\otimes n}_2 U_{\text{QFM}}^\dagger(\boldsymbol{x},\boldsymbol{\theta}_K) U_{\text{QFM}}(\boldsymbol{x},\boldsymbol{\theta}_Q) \ket{0}^{\otimes n}_2)}{2}  \\
&= \frac{1 - \text{Im}(\braket{K|Q})}{2}. 
\end{align*} 
\end{itemize}

The improved Hadamard test circuit effectively extracts both the real and imaginary components of the quantum self-attention weights.

\subsection{Quantum Complex Weight Embedding Module}
In the previous section, we derived the real and imaginary components of the complex self-attention weights $\braket{K|Q}$. To incorporate these weights as coefficients in the Complex Linear Combination of Unitaries (CLCUs) for subsequent quantum computations, we convert them into amplitude and phase representations for encoding into the quantum circuit. We achieve this using a block encoding technique based on Fast Approximate Quantum Circuits for Block-Encodings (FABLE) \cite{camps2022fable}, which we have optimized and simplified to enable efficient encoding of complex attention weights.

In our approach, we embed the complex attention weight matrix into the diagonal subspace of an expanded unitary matrix, ensuring the accurate representation of both the magnitude and phase of the complex information in the computational basis. Through post-selection techniques, we then extract the state of the target qubit system, which can be used for subsequent quantum operations. Specifically, we represent the real and imaginary components of the complex coefficients as their amplitude and phase form: 
\begin{equation}
    \braket{K|Q} = \text{Re}(\braket{K|Q})+\mathrm{i}\text{Im}(\braket{K|Q})=|\braket{K|Q}| e^{\mathrm{i} \phi}, 
\end{equation}
where 
\begin{equation}
|\braket{K|Q}| = \sqrt{\big(\text{Re}(\braket{K|Q})\big)^2 + \big(\text{Im}(\braket{K|Q})\big)^2}, 
\end{equation}
and 
\begin{equation}
\phi = \arctan\!\left(\frac{\text{Im}(\braket{K|Q})}{\text{Re}(\braket{K|Q})}\right), 
\end{equation}
$|\braket{K|Q}|$ is the magnitude and $\phi$ is the phase. This transformation facilitates embedding the complex information into the block encoding framework.

The entire encoding process proceeds as follows:
\begin{equation} 
\begin{aligned} 
    &\ket{0}\ket{0}^{\otimes n} \ \xrightarrow{H^{\otimes n}} \frac{1}{\sqrt{2^{n}}}\sum_{j=0}^{2^{n}-1}\ket{0}\ket{j} \\
    &\xrightarrow{O_A} \frac{1}{\sqrt{2^{n}}}\sum_{j=0}^{2^{n}-1}\left( \cos(\theta_{ij})e^{-i\phi_{j}}\ket{0} + \sin(\theta_{j})e^{\mathrm{i}\phi_{j}}\right)\ket{j},  \\
    &\xrightarrow{P_0} \frac{1}{\mathcal{N}}\sum_{j=0}^{2^{n}-1}\cos(\theta_{j})e^{-\mathrm{i}\phi_{ij}}\ket{j},
\end{aligned} 
\end{equation}
where $n$ represents the number of working qubits. $P_0$ refers to the post-selection measurement, where the highest bit collapses to $\ket{0}$, thereby extracting the state of the remaining qubit system and normalizing it. $O_A$ denotes the block encoding operation, which consists of all controlled $R_y$ and $R_z$ gates.

The $\cos(\theta_{ij})$ term, representing the magnitude of the attention weight, is implemented using the controlled $R_y(\theta_{ij})$ gate, which adjusts the amplitude of the quantum state. Meanwhile, the phase shift, $e^{i\phi_{ij}}$, is realized using the controlled $R_z(\phi_{ij})$ gate, which introduces the desired phase. By combining these two controlled gates, both the magnitude and phase information of the attention weights are accurately encoded into the quantum state.

\begin{figure}[h]
  \centering
  \includegraphics[width=3.3in]{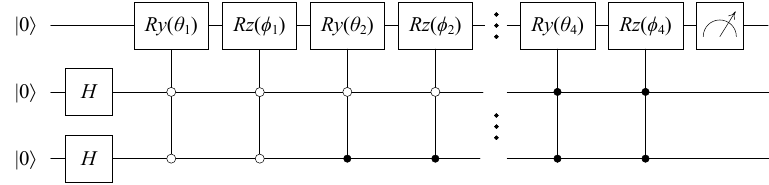}
  \caption{The architecture of 3 qubits circuit for encoding quantum attention weights.}
  \label{Circuit for Encoding Attention Weights}
  \end{figure}

To provide an intuitive illustration of the encoding process, we consider a specific example using a 3 qubits quantum circuit to encode complex values into the computational basis, as shown in Fig. \ref{Circuit for Encoding Attention Weights}. We utilize controlled $C^2(R_y(\theta_i))$ and $C^2(R_z(\phi_i))$ gates on auxiliary qubits to precisely embed the magnitude, $\cos(\theta_i)$, and phase, $e^{-i\phi_i}$, of the complex attention weights into the diagonal elements of a $4 \times 4$ submatrix of the larger $8 \times 8$ unitary matrix $O_A$.

\begin{equation}
\begin{array}{@{}l@{}}
  O_{A} = \\[0.5em]
  \begingroup
  \setlength{\arraycolsep}{-15pt} 
  \begin{bmatrix}
      \cos(\theta_{0})e^{-\mathrm{i}\phi_{0}} & & & & -\sin{\theta_{0}}e^{-\mathrm{i}\phi_{0}} & & & \\
      & \cos(\theta_{1})e^{-\mathrm{i}\phi_{1}} & & & & -\sin{\theta_{1}}e^{-\mathrm{i}\phi_{1}} & & \\
      & & \cos(\theta_{2})e^{-\mathrm{i}\phi_{2}} & & & & -\sin{\theta_{2}}e^{-\mathrm{i}\phi_{2}} & \\
      & & & \cos(\theta_{3})e^{-\mathrm{i}\phi_{3}} & & & & -\sin{\theta_{3}}e^{-\mathrm{i}\phi_{3}} \\
      \sin{\theta_{0}}e^{\mathrm{i}\phi_{0}} & & & & \cos(\theta_{0})e^{\mathrm{i}\phi_{0}} & & & \\
      & \sin{\theta_{1}}e^{\mathrm{i}\phi_{1}} & & & & \cos{\theta_{1}}e^{\mathrm{i}\phi_{1}} & & \\
      & & \sin{\theta_{2}}e^{\mathrm{i}\phi_{2}} & & & & \cos{\theta_{2}}e^{\mathrm{i}\phi_{2}} & \\
      & & & \sin{\theta_{3}}e^{\mathrm{i}\phi_{3}} & & & & \cos{\theta_{3}}e^{\mathrm{i}\phi_{3}}.
  \end{bmatrix}
  \endgroup
\end{array}
\end{equation}

First, we apply a Hadamard gate to create a uniform superposition state. Then, we encode the matrix using the OA technique. Afterward, we perform a post-selection measurement on the highest qubit, retaining only the cases where the measurement result is $\ket{0}$. This ensures the system's state is projected onto the desired $4 \times 4$ subspace, with the complex coefficients encoded into the computational basis of the remaining two qubits:
\begin{equation}
  \begin{aligned}
  \ket{\psi} &= \cos(\theta_0)e^{-\mathrm{i}\phi_0}\ket{00} + \cos(\theta_1)e^{-\mathrm{i}\phi_1}\ket{01} \\
  &\quad + \cos(\theta_2)e^{-\mathrm{i}\phi_2}\ket{10} + \cos(\theta_3)e^{-\mathrm{i}\phi_3}\ket{11}.
  \end{aligned}
\end{equation}

In this final state, the two remaining working qubits, in the computational basis, precisely reflect the magnitude and phase information of the original $4 \times 4$ complex submatrix.

\subsection{Quantum Complex Linear Combination of Unitaries}
The LCUs method implements specific quantum operations by linearly combining multiple unitary operators with real coefficients. While effective, this approach encounters limitations when dealing with the complex nature of quantum states. Since the coefficients in LCUs are constrained to real values, any representation of complex effects or phase information must be indirectly encoded by designing the unitary operators $U_j$ to incorporate additional quantum gates that encode the phase (via $\alpha_jU_j = |\alpha_j|e^{i\phi_j}U_j = |\alpha_j|U'_j$). This added complexity not only increases the implementation difficulty but also limits the flexibility in choosing unitary operators. In contrast, our CLCUs method directly utilizes complex coefficients, introducing an assumption that is more aligned with the inherent nature of quantum mechanics. This design of CLCU reduces the dependency on the structure of unitary operators, allowing the model to directly adjust complex coefficients during optimization to capture the relationships between quantum states. This inductive bias makes CLCU more adept at efficiently learning tasks based on the inner product weight calculations in quantum self-attention mechanisms.

\begin{figure}[h]
  \centering
  \includegraphics[width=3.5in]{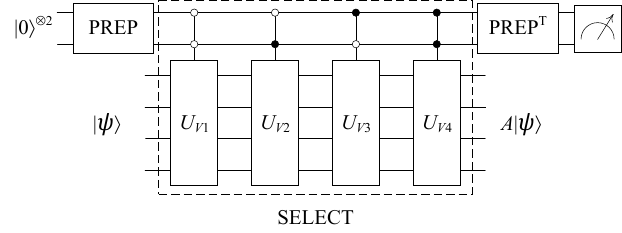}
  \caption{The architecture of CLCUs for implementing $A\ket{\psi} = \frac{1}{\mathcal{N}'}(\alpha_1U_{V1} + \alpha_2U_{V2} + \alpha_3U_{V3} + \alpha_4U_{V4})\ket{\psi}$}
  \label{Circuit for CLCUs}
  \end{figure}

\begin{definition}[Quantum Complex Linear Combination of Unitaries, CLCUs] Given a set of unitary operators ${U_1, U_2, ..., U_N}$, with corresponding complex coefficients ${\alpha_1, \alpha_2, ..., \alpha_N}$ (where $\alpha_j \in \mathbb{C}, \alpha_j = |\alpha_j| e^{\mathrm{i}\theta_j}$), the CLCUs operator $A$ acts on a quantum state $\ket{\psi}$ as follows: 
\begin{equation} 
  A \ket{\psi} = \frac{1}{\mathcal{N}'}\sum_{j=0}^{N-1} \alpha_j U_j \ket{\psi} = \frac{1}{\mathcal{N}'}\sum_{j=0}^{N-1} |\alpha_j| e^{\mathrm{i} \theta_j} U_j \ket{\psi}, 
\end{equation} 
\end{definition}
where $\mathcal{N}' = || \sum_{j=0}^{N-1} \alpha_j U_j \ket{\psi} ||$ is the normalization constant.

The CLCUs implement a linear combination of unitary operations with complex coefficients using auxiliary qubits, conditional operations, and post-selection measurements. The specific implementation steps are outlined below and depicted in Fig. \ref{Circuit for CLCUs}.

First, during the preparation operation $U_{\text{PREP}}$, phase information is introduced to encode the complex coefficients. Specifically, we prepare:
\begin{equation}
  U_{\text{PREP}} \ket{0}^{\otimes n} = \frac{1}{\sqrt{\mathcal{N}}} \sum_{j=0}^{N-1} \sqrt{|\alpha_j|} e^{\mathrm{i} \theta_j / 2} \ket{j},
\end{equation}
where $ \mathcal{N} = \sum_{j=0}^{N-1} |\alpha_j|$ is the normalization constant, and the introduction of $\theta_j/2$  ensures that the phase $\theta_k$ can accumulate correctly in subsequent operations.

The selection operator $U_{\text{SELECT}}$ applies the corresponding unitary $U_k$ to the target state $\ket{\psi}$ conditioned on the auxiliary state $\ket{0}$:
\begin{equation}
  U_{\text{SELECT}} \ket{j} \ket{\psi} = \ket{j} U_j \ket{\psi}.
\end{equation}

Unlike the standard LCUs method, which uses $U_{\text{PREP}}^\dagger$, here we employ the transpose $U_{\text{PREP}}^T$. Assuming the modified preparation operation $U_{\text{PREP}}$ is expressed as:
\begin{equation}
  U_{\text{PREP}} = \frac{1}{\sqrt{\mathcal{N}}}{\sum_{j=0}^{N-1} \sqrt{|\alpha_j|} e^{\mathrm{i} \theta_j / 2} \ket{j} \bra{0}^{\otimes n} + \text{orthogonal terms}},
\end{equation}
its transpose form is:
\begin{equation}
  U_{\text{PREP}}^T = \frac{1}{\sqrt{\mathcal{N}}} \sum_{j=0}^{N-1} \sqrt{|\alpha_j|} e^{\mathrm{i} \theta_j / 2} \ket{0}^{\otimes n} \bra{j} + \text{orthogonal terms}.
\end{equation}

Combining the above operations, we derive the effect of the combined operations in detail. First, prepare the auxiliary state and apply the selection operation to get:
\begin{equation}
  \begin{aligned}
    &U_{\text{SELECT}} (U_{\text{PREP}}\otimes I_{\text{target}}) \ket{0}^{\otimes n} \ket{\psi} \\
    = &\frac{1}{\sqrt{\mathcal{N}}} \sum_{j=0}^{N-1} \sqrt{|\alpha_j|} e^{\mathrm{i} \theta_j / 2} \ket{j} U_j \ket{\psi}.
\end{aligned}
\end{equation}

Then, apply the transpose preparation operation $U_{\text{PREP}}^T$ to get:
\begin{equation}
\begin{split}
  &(U_{\text{PREP}}^T \otimes I_{\text{target}}) U_{\text{SELECT}} (U_{\text{PREP}} \otimes I_{\text{target}}) \ket{0}^{\otimes n} \ket{\psi}\\
  = &\frac{1}{\mathcal{N}} \sum_{j=0}^{N-1} |\alpha_j| e^{\mathrm{i} \theta_j} \ket{0}^{\otimes n} U_j \ket{\psi} + \text{orthogonal terms}.
\end{split}
\end{equation}

Finally, by measuring the auxiliary qubit, if the measurement result is $\ket{0}$, the target state is projected to:
\begin{equation}
  A \ket{\psi} =\frac{1}{\mathcal{N}'} \sum_{j=0}^{N-1} \alpha_j U_j \ket{\psi}.
\end{equation}

To ensure the effectiveness and feasibility of the method, the following key points should be considered:
\begin{itemize}
  \item When introducing phases in the preparation operation, precise control of the phases is required to avoid error accumulation.
  \item It is important to ensure that both $U_{\text{PREP}}$ and its transpose (non-conjugate) operation $U_{\text{PREP}}^T$ are unitary transformations. This guarantees their reversibility and the physical feasibility of their implementation in quantum computation. In practice, $U_{\text{PREP}}$ is typically constructed using gates such as $H$, $CR_z(\theta)$ and $CR_y(\theta)$, with their transposed gates being $H$, $CR_z(\theta)$ and $CR_y(-\theta)$. These gates preserve unitarity and can be directly implemented in quantum circuits.
\end{itemize}

With these modifications, the LCUs method is successfully extended to handle linear combination operations with complex coefficients.

In this paper, CLCUs enhance our quantum self-attention model through three applications: Quantum Similarity-Driven Complex Weighted Sum; Trainable Complex Weighted Sum; Quantum Multi-Head Self-Attention Mechanism.

\textbf{Quantum Similarity-Driven Complex Weighted Sum}: In the quantum self-attention mechanism, the attention weights determine the contribution of each quantum state to the final representation. Consider a set of quantum states ${\{\ket{U_j}\}_{j=0}^{N-1}}$ and their corresponding attention weights ${\{\alpha_j}\}_{j=0}^{N-1}$. We aim to implement a quantum state representation as a weighted sum using a quantum circuit. The attention weight encoding module leverages the CLCUs method to encode each attention weight $\braket{K_j|Q_k}$ into the corresponding quantum circuit $U_{Vj}$:
\begin{equation} 
\begin{split}
      \ket{S_k} &= \frac{1}{\mathcal{N}_{S_i}}\sum_{j=0}^{N-1} \braket{K_j|Q_k} U_{\text{QFM}}(\boldsymbol{x},\boldsymbol{\theta}_{V_j}) \ket{0}^{\otimes n} \\
      &= \frac{1}{\mathcal{N}_{S_i}}\sum_{j=0}^{N-1} \braket{K_j|Q_k}\ket{V_j},
\end{split}
\end{equation} 
where $\braket{K_j|Q_k}$ represents the inner product between the quantum states $\ket{K_j} $ and $\ket{Q_k}$.

\textbf{Trainable Complex Weighted Sum}: After the attention weights have been encoded, we introduce the weighted sum module, which uses independent CLCUs operations to perform a weighted sum on the generated weighted quantum states $\{\ket{S_j}\}_{j=0}^{M-1}$, forming the global quantum state $\ket{G}$:
\begin{equation} 
  \ket{G} = \frac{1}{\mathcal{N}_{G}}\sum_{j=0}^{M-1} \beta_j \ket{S_j},
\end{equation}
where $\beta_j$ represents the trainable complex weight of the $j$-th quantum state $\ket{S_j}$.The magnitudes and phases of these coefficients can be dynamically adjusted through parameterized quantum gates.

\textbf{Quantum Multi-Head Self-Attention Mechanism}: To further improve the expressiveness of the quantum self-attention mechanism, we introduce the Multi-Head Attention mechanism. In the classical Transformer model, multi-head self-attention captures different feature representations through parallel self-attention heads, boosting the model’s capability. We adopt a similar approach in the quantum self-attention mechanism by implementing multi-head self-attention using multiple independent CLCUs operations.

Specifically, consider $H$ self-attention heads, each with its own set of attention weights $\{\gamma^{(h)}\}_{h=0}^{H-1}$ and corresponding quantum states $\{\ket{G^{(h)}}\}_{h=0}^{H-1}$. Then, through CLCUs operations, we weight and sum all $\ket{G^{(h)}}$ states to form the final global quantum state:
\begin{equation}
  \ket{\psi_{\text{final}}} = \frac{1}{\mathcal{N}_{\text{final}}}\sum_{h=0}^{H-1} \gamma^{(h)}\ket{G^{(h)}},
\end{equation}
where $\gamma_h$ represents the trainable global complex coefficients, which are encoded through independent CLCUs operations.

\subsection{Quantum Feedforward Neural Network}
In the classic Transformer architecture, the Feed-Forward Network (FFN) layer performs feature transformations to enhance the model's ability to process. Similarly, to improve the expressiveness and flexibility of the quantum self-attention mechanism, we introduce a trainable quantum circuit layer within the quantum self-attention framework. This layer increases the complexity and entanglement of quantum states, thus boosting the expressiveness of the quantum self-attention mechanism.

We adopt a hardware-efficient quantum circuit layer \cite{du2023problem}, which consists of a sequence of trainable $R_z$ and $R_y$ rotation gates followed by CNOT gates for entanglement. The circuit structure is defined as:
\begin{equation}
  U_l(\boldsymbol{\theta}) = \bigotimes_{j=1}^n \left( R_z(\theta^{(l,j,1)}) \, R_y(\theta^{(l,j,2)}) \, R_z(\theta^{(l,j,3)}) \right) U_{\text{ent}},
\end{equation}
where $U_{\text{ent}}$ is the entanglement layer, formed by CNOT gates, used to introduce entanglement between qubits. $l$ represents the layer number. $j$ represents the qubit index.

For details on the architecture of the QFFN, please refer to the supplementary file \ref{section: Circuit Architecture}.

\subsection{Loss Function}
In this paper, we focus on classification tasks with 2, 3, and 4 classes. For binary classification tasks, the measurement strategy is simplified to measuring only the first qubit in the $\sigma_z$ basis. As the number of categories increases to three, measurements are taken in the $\sigma_x$, $\sigma_y$, and $\sigma_z$ bases for a three-category task. For multi-class tasks, we adopt a tensor-product measurement strategy across two qubits, generating multidimensional expectation values to support up to nine classes. This approach ensures sufficient independent observables as the task complexity increases.
\begin{equation}
\begin{split}
M_j = 
\begin{cases}
(-1)^j\sigma_z^{(0)} & \text{if } n=2, \, j \in \{0, 1\} \\[.5em]
\sigma_{p(j)}^{(0)} & \text{if } n=3, \, j \in \{0, 1, 2\} \\[.5em]
\sigma_{p(j \bmod 3)}^{(0)} \otimes \sigma_{p(\lfloor j/3 \rfloor \bmod 3)}^{(1)} & \begin{aligned}
& \text{if } 3<n\leq9, \\
& j \in \{0, 1, \dots, n-1\}
\end{aligned}
\end{cases}
\end{split}
\end{equation}
where $\sigma^{(k)}_p$ denotes the Pauli operator acting on the $k$-th qubit. $p(i)$ is a function mapping an index $i \in \{0, 1, 2\}$ to a Pauli operator basis: $p(0) = x$; $p(1) = y$; $p(2) = z$. That is, $\sigma_{p(0)} = \sigma_x$, $\sigma_{p(1)} = \sigma_y$, and $\sigma_{p(2)} = \sigma_z$. $\lfloor \cdot \rfloor$ denotes the floor function. $\bmod$ denotes the modulo operation. For instance, when $n=4$, the operators are $\sigma^{(0)}_{x} \otimes \sigma^{(1)}_{x}$, $\sigma^{(0)}_{y} \otimes \sigma^{(1)}_{x}$, $\sigma^{(0)}_{z} \otimes \sigma^{(1)}_{x}$, $\sigma^{(0)}_{x} \otimes \sigma^{(1)}_{y}$.

The resulting probability distribution is given by:
\begin{equation}
  \hat{y}_k = \frac{1 + \langle\psi|M_k|\psi\rangle}{\sum_{j=0}^{n-1} {(1 + \langle\psi|M_j|\psi\rangle})}, \quad k\in\{0,1,2,...,n-1\}.
\end{equation}

The loss function is computed using the simple cross-entropy formula:
\begin{equation}
  \mathcal{L} = -\frac{1}{N} \sum_{j=0}^{N-1} \sum_{c=0}^{C-1} y_{j,c} \log(\hat{y}_{j,c}),
\end{equation}
where $N$ is the total number of samples in the training dataset. $C$ is the number of categories in the classification task. $y_{j,c}$ is the true label of sample $j$ for category $c$. $\hat{y}_{j,c}$ is the predicted probability that sample $j$ belongs to category $c$.

\section{Numerical Experiments}
In this study, we evaluate the performance of our proposed quantum self-attention mechanism through numerical simulations using two widely recognized image classification datasets, MNIST and Fashion-MNIST. In our primary benchmarking experiments, we compare our model against three quantum self-attention models: QKSAN, QSAN, and GQHAN. All models are trained and tested under identical conditions with consistent training and test set sizes, ensuring a fair and direct comparison of their performance.

We further extend our evaluation by exploring the scalability of our approach across both task complexity and quantum system size. Specifically, we conduct experiments on 2, 3, and 4 class classification tasks as well as on quantum systems ranging from 3-qubit to 8-qubit configurations. These extension experiments provide insights into how our quantum self-attention mechanism adapts to larger, more complex quantum architectures and handles more challenging classification scenarios. Furthermore, we conduct ablation studies to compare models utilizing complex-valued self-attention weights against those employing real-valued weights. 

\subsection{Experimental Setup}
In the data preprocessing stage, we begin by dividing the raw images into patches. We then apply PCA to reduce the dimensionality of the features in each image patch, aligning it with the number of qubits in the quantum model. To minimize the influence of preprocessing on the experimental results, we deliberately use a non-trainable, fixed-parameter PCA for dimensionality reduction. This approach, based on linear transformations, is simple, and introduces no additional learnable parameters, ensuring that any differences in classification performance are primarily due to the quantum self-attention mechanism, rather than the preprocessing techniques. Additionally, to ensure better alignment with quantum state representations, we normalize all input data to the range $[0, \pi]$.

For implementation, we use the TensorCircuit \cite{zhang2023tensorcircuit} framework to simulate the quantum circuits, integrating it with TensorFlow \cite{abadi2015tensorflow} for parameter optimization. The Adam optimizer \cite{kingma2014adam} is employed with a batch size of 32. In both the MNIST and Fashion-MNIST datasets, we randomly select 512 samples per class from the training set and 128 samples per class from the test set. For the quantum single-head self-attention model, we divide each image into 4 patches, while for the quantum dual-head self-attention model, one set of images is divided into 4 patches, and the other into 49 patches. Each experiment is repeated 5 times using different random seeds, and the final results are averaged to ensure robustness and reduce variability.

\subsection{Comparison with Existing Quantum Self-Attention Models}
In this section, we compare our model with three quantum self-attention models: QKSAN, QSAN, and GQHAN. The evaluation is performed under the same experimental conditions, with each model trained using 50 samples per class and tested on 500 samples per class.

\begin{table}[h]
  \centering
  \caption{Performance Comparison on MNIST Dataset}
  \begin{tabular}{|l|c|c|c|l}
  \hline
  \textbf{Model}         & \textbf{Test Accuracy} & \textbf{Train Accuracy} & \textbf{Qubits} \\ \hline
  Ours                   & 100\%                & 100\%                & 4               \\ \hline
  QKSAN \cite{zhao2024qksan}                & 99.0\%               & 99.06\%              & 4            \\ \hline
  QSAN \cite{shi2024qsan}                & 100\%                & 100\%                & 8             \\ \hline
  \end{tabular}
  \label{table: compare to QSAN, QKSAN on MNIST Dataset}
\end{table}
  
\begin{table}[h]
  \centering
  \caption{Performance Comparison on Fashion MNIST Dataset}
  \begin{tabular}{|l|c|c|c|}
  \hline
  \textbf{Model}         & \textbf{Test Accuracy}  & \textbf{Train Accuracy}  & \textbf{Qubits}    \\ \hline
  Ours            & 99.2±0.7483 \%   & 98.4±0.5514\%      & 4                \\ \hline
  QKSAN \cite{zhao2024qksan}               &  98\%          & 97.22\%            & 4                \\ \hline
  QSAN \cite{shi2024qsan}                & 96.8\%           & 96.77\%            & 8                \\ \hline
  GQHAN \cite{zhao2024gqhan}               & 98.59\%           & 98.65\%            & 4               \\ \hline
  \end{tabular}
  \label{table: compare to QSAN, QKSAN, GQHAN on Fashion-MNIST Dataset}
\end{table}

On the MNIST dataset, as shown in Table \ref{table: compare to QSAN, QKSAN on MNIST Dataset}, our model demonstrates a significant performance advantage. Using only 4 qubits, our approach achieves 100\% accuracy on both the training and test sets, a level of performance unmatched by competing models. Our model outperforms QKSAN, QSAN and GQHAN in the Fashion-MNIST classification task shown in Table \ref{table: compare to QSAN, QKSAN, GQHAN on Fashion-MNIST Dataset} by achieving higher average accuracy with fewer qubits.

We attribute the breakthrough performance of our model, particularly under small sample sizes and low qubit counts, to the innovative design of its quantum state similarity measure. Specifically, QKSAN employs a quantum kernel method to compute the similarity between quantum states $\ket{Q}$ and $\ket{K}$ by evaluating the magnitude of their inner product, which yields a real-valued result. QSAN, in contrast, uses a CNOT gate-based strategy to integrate the quantum states $\ket{Q}$ and $\ket{K}$, directly fusing them for self-attention calculations. Although this direct integration is straightforward, it does not capture the subtle nuances and intricate relationships between quantum states. GQHAN eschews a theoretical similarity measure altogether, instead relying on a flexible "Oracle" mechanism to weight the data without offering a quantifiable assessment of state similarity. In contrast, our model incorporates an improved Hadamard test that measures both the real and imaginary components of the quantum state similarity, thereby fully capturing the phase information to quantum states.

\subsection{Scalability Analysis of Model Performance}
In this section, we analyze the scalability of the proposed quantum self-attention mechanism under different experimental setups. Specifically, we explore the impact of the number of classification tasks (2, 3, and 4 class) on model performance, the effects of quantum single-head and dual-head self-attention mechanisms, the influence of the number of qubits (ranging from 3 to 8) on model performance and training stability, as well as the effect of dataset size on the model's generalization ability.

\begin{figure}[h]
  \centering
  \includegraphics[width=2.8in]{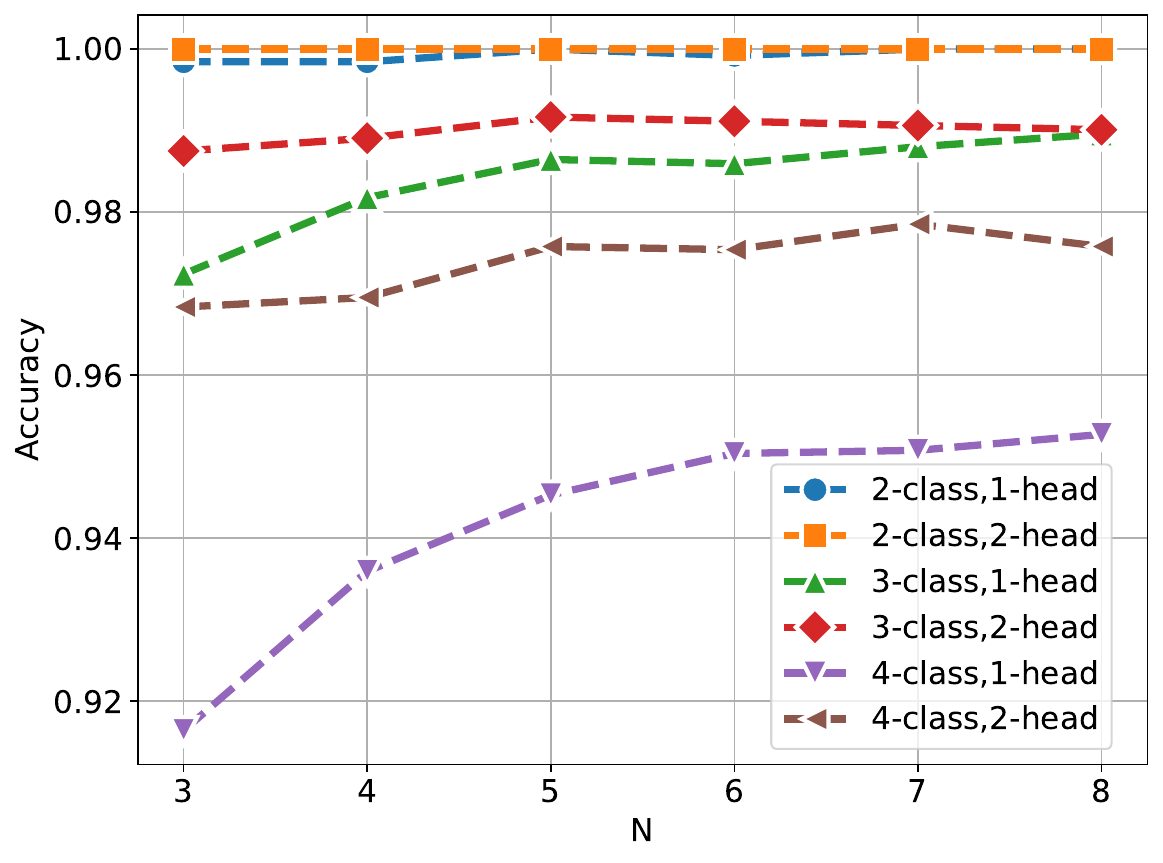}
  \caption{Scalability of Our Models on MNIST with Varying Qubits, Classification Tasks, and Multi-Head Attention.}
  \label{fig: Scalability on MNIST}
  \end{figure}

\begin{figure}[h]
  \centering
  \includegraphics[width=2.8in]{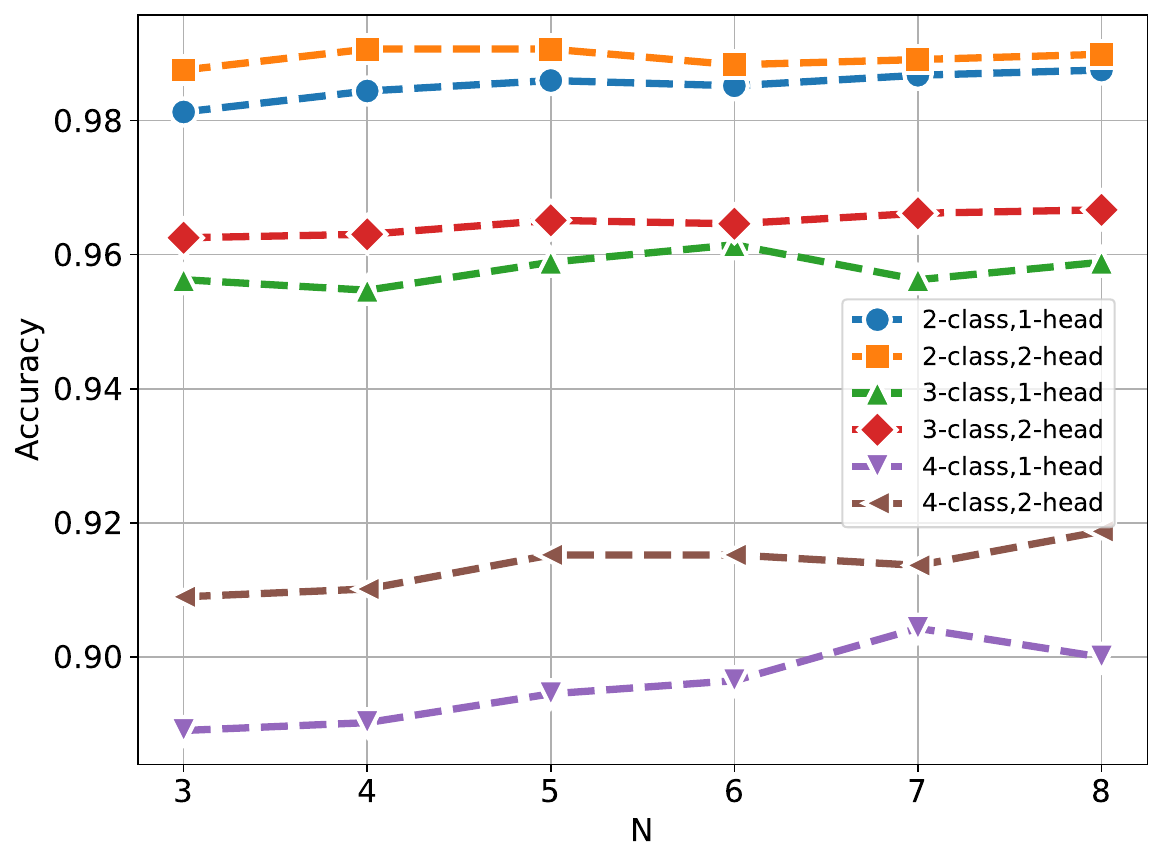}
  \caption{Scalability of Our Models on Fashion MNIST with Varying Qubits, Classification Tasks, and Multi-Head Attention.}
  \label{fig: Scalability on Fashion MNIST}
  \end{figure}

Our experimental results, shown in Fig. \ref{fig: Scalability on MNIST} and \ref{fig: Scalability on Fashion MNIST}, highlight several key trends. Regarding the impact of the number of classification tasks on performance, for 2-class classification tasks, the results are nearly perfect, especially on MNIST where test accuracy with 3 to 5 qubits often approaches or reaches 100\%. This suggests that such binary tasks are relatively straightforward for the quantum self-attention mechanism. Predictably, overall test accuracy decreased as task complexity rose from 2-class to 4-class classification. However, for these more complex 3- and 4-class tasks, model performance generally improved with an increasing number of qubits. For instance, on MNIST, the single-head 3-class accuracy rose from 97.24\% (3 qubits) to nearly 99\% (8 qubits), while on Fashion-MNIST, it increased from 95.63\% to 96.61\%. This indicates that augmenting the number of qubits enhances the model's representational capacity, potentially leading to better performance on more challenging tasks by allowing it to model more complex data relationships.

Regarding the impact of multi-head quantum self-attention on performance, dual-head architectures consistently demonstrate superior performance metrics compared to single-head configurations across all classification tasks. This improvement stems from the multi-head mechanism's ability to introduce independent attention heads, which can capture diverse feature representations from different input subspaces, thereby enhancing the model's expressiveness and classification accuracy. In essence, the additional parameters inherent in the quantum multi-head design effectively contribute to boosting model performance for these tasks. However, for simpler tasks such as binary and ternary classification in MNIST, the performance gap between single-head and multi-head mechanisms narrows as the number of qubits increases (particularly beyond 6 qubits). This suggests that in these simpler tasks, the single-head attention mechanism already has sufficient expressive power, and the advantages of the multi-head mechanism diminish.

\begin{figure}[h]
  \centering
  \includegraphics[width=2.8in]{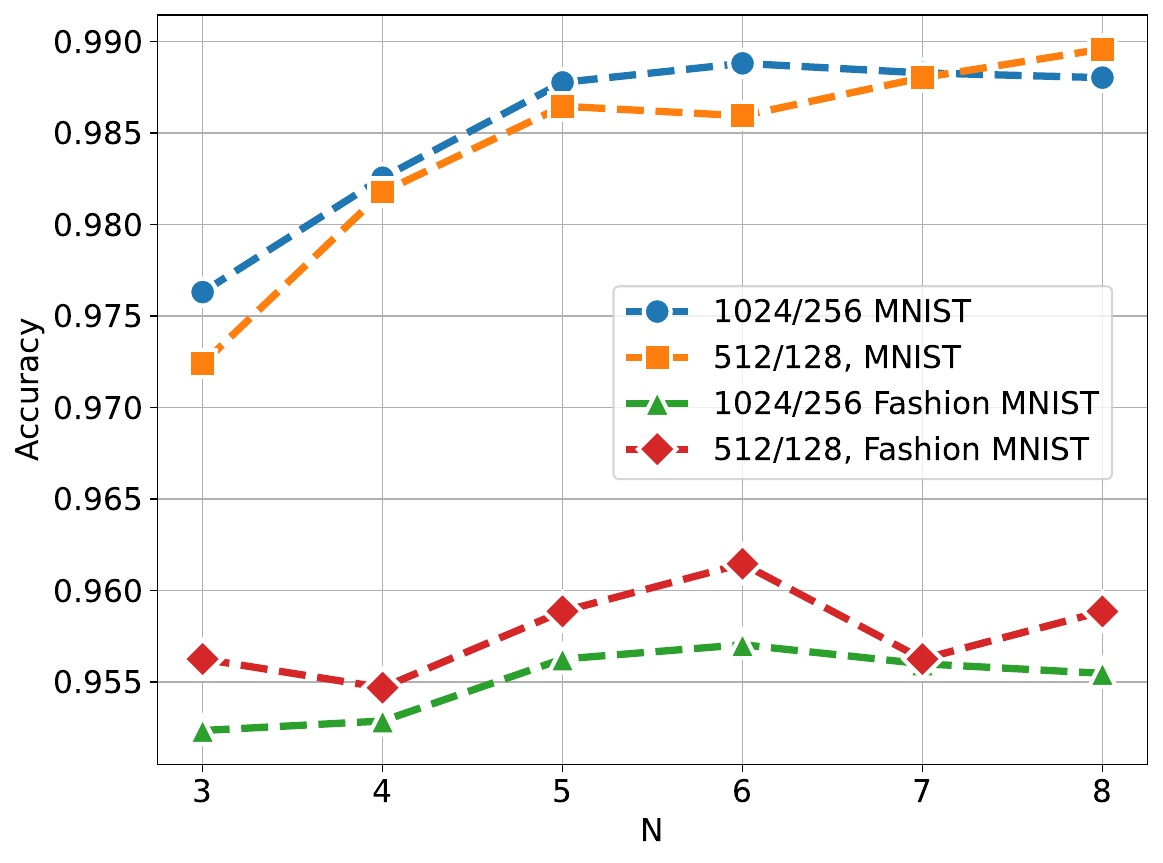}
  \caption{Performance Comparison of 3-Class with Varying Qubits and Dataset Sizes.}
  \label{fig: 1024 sizes}
  \end{figure} 

In our experiments, when using a small-scale training set, we found that as the number of qubits increased, the test accuracy generally showed an upward trend, with some local fluctuations. For example, as seen in the table, for the ternary and quaternary tasks, when the number of qubits increased from 3 to 5, the test accuracy steadily improved. However, at 6 qubits, some experiments (such as the 4-class task on Fashion-MNIST) showed a slight decline. We believe that these local fluctuations might be due to a small training data size, data noise, or sample randomness. To further validate this hypothesis, we expanded the training data to 1,024 samples and the test data to 256 samples. As shown in Fig. \ref{fig: 1024 sizes}, the results indicated that under larger dataset conditions, the test accuracy followed a pattern of first increasing and then slightly decreasing. A moderate increase in the number of qubits enhanced the model's feature representation ability, helping capture more data features. However, when the model complexity, which generally increases with the number of qubits, exceeds a certain threshold, overfitting emerges, impairing the model's generalization ability. This phenomenon is consistent with the conclusions of Ref. \cite{du2023problem}. They found that the expected risk decreases and then increases as the model complexity increases, exhibiting a U-shaped behavior.

\subsection{Ablation Study on the Impact of Quantum Self-Attention Weights}
\label{section: Ablation}
In this section, we analyze the impact of two quantum self-attention weight calculation methods on model performance through an ablation study: One method is based on real-valued overlap quantum similarity calculations, utilizing strategies such as the quantum kernel function and the SWAP test to compute the similarity between two quantum states. The other using our proposed improved Hadamard test method. Specifically, when the quantum inner product calculation uses real-valued overlap quantum similarity, the corresponding LCUs coefficients are real numbers; while when it uses complex-valued overlap quantum similarity, CLCUs coefficients are complex. We evaluated the performance differences of these two methods under conditions with 3 to 8 qubits for the 3-class classification tasks on MNIST and Fashion-MNIST.

\begin{figure}[h]
  \centering
  \includegraphics[width=2.8in]{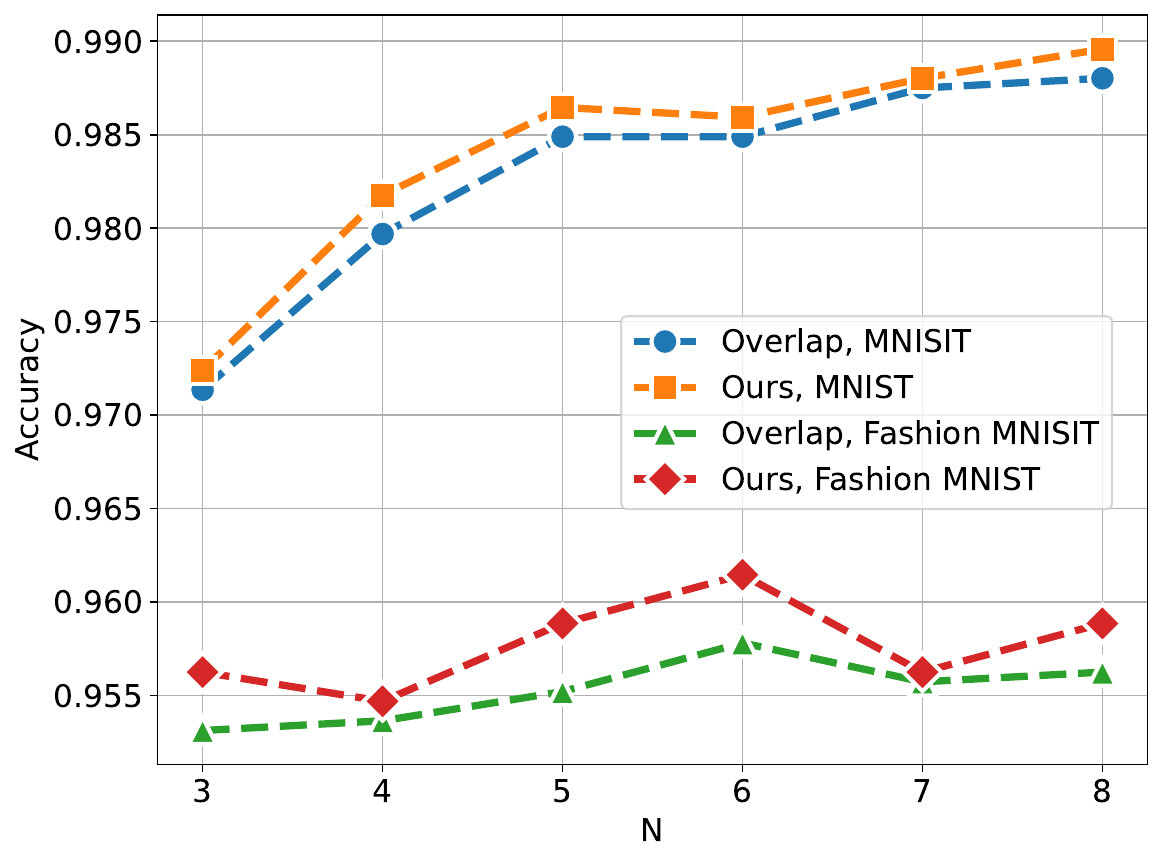}
  \caption{Performance Comparison of Quantum Attention Weight Methods.}
  \label{fig: ablation study}
  \end{figure}

Fig. \ref{fig: ablation study} shows that the improved Hadamard test method (Ours) outperforms the method based on real-valued overlap quantum similarity in terms of average accuracy across all qubit configurations for both the MNIST and Fashion-MNIST datasets. Our method captures the complex similarity between quantum states, while the SWAP test and quantum kernel function methods only consider the real-valued overlap quantum similarity. This additional phase information, under experimental conditions with small samples and limited resources, demonstrates more flexible and efficient quantum state representation, enabling the model to make better use of the limited quantum resources and enhancing the expressive power of the quantum self-attention mechanism.

\section{Conclusion}
In this paper, we introduce a novel quantum self-attention mechanism that integrates both amplitude and phase information in its attention weights, extending the classical self-attention framework into the quantum domain. By leveraging complex-valued attention weights, our approach provides a more expressive representation of quantum states, allowing the model to better distinguish and utilize complex input patterns for improved performance.

Through extensive experimental validation, we demonstrated that QCSAM outperforms current quantum self-attention models, including QKSAN, QSAN, and GQHAN, in terms of both classification accuracy and efficiency. The use of complex-valued quantum attention weights significantly enhances the model's ability to capture subtle dependencies in quantum data, even with limited qubit resources. Moreover, the integration of multi-head attention further boosts the model's representational capacity, allowing for more effective utilization of quantum states in classification tasks.

In future work, we aim to explore the fundamental differences between classical and quantum self-attention mechanisms, focusing on how quantum properties such as superposition and entanglement influence the attention process. This will allow for a deeper understanding of the advantages and challenges of integrating quantum techniques into classical models, ultimately guiding the development of more efficient and powerful quantum machine learning algorithms.

%

\appendices

\ifCLASSOPTIONcompsoc
  \section*{Acknowledgments}
\else
  \section*{Acknowledgment}
\fi

This should be a simple paragraph before the References to thank those individuals and institutions who have supported your work on this article.

\ifCLASSOPTIONcaptionsoff
  \newpage
\fi

\bibliographystyle{IEEEtran} 
\bibliography{references}    

\newpage
\onecolumn

\begin{center}
    \Large \textbf{Supplementary File}
\end{center}

This supplementary file provides additional technical details of our proposed models. In the following sections, we present comprehensive experimental results and a detailed description of the circuit architectures. The Appendix \ref{section: Detailed Experimental Results} reports extensive performance metrics on the MNIST and Fashion-MNIST datasets under various settings, while the the Appendix \ref{section: Circuit Architecture} explains the design and implementation of our quantum state embedding circuit and the QFFN.

\section{Detailed Experimental Results}
\label{section: Detailed Experimental Results}

\begin{table}[h]
\centering
\caption{Results on MNIST and Fashion-MNIST Datasets (Test set: 512, Training set: 128)}
\footnotesize  
\setlength{\tabcolsep}{2pt}  
\begin{tabular}{c|ccc|ccc|ccc|ccc}
\hline
\multirow{3}{*}{Qubits} & \multicolumn{6}{c|}{MNIST} & \multicolumn{6}{c}{Fashion-MNIST} \\
\cline{2-13}
& \multicolumn{3}{c|}{1H} & \multicolumn{3}{c|}{2H} & \multicolumn{3}{c|}{1H} & \multicolumn{3}{c}{2H} \\
\cline{2-13}
& 2-class & 3-class & 4-class & 2-class & 3-class & 4-class & 2-class & 3-class & 4-class & 2-class & 3-class & 4-class \\
\hline
3 & 99.84$\pm$0.19 & 97.24$\pm$0.35 & 91.64$\pm$0.67 & 100.00$\pm$0.00 & 98.75$\pm$0.51 & 96.84$\pm$0.45 & 98.13$\pm$0.97 & 95.63$\pm$0.78 & 88.91$\pm$1.34 & 98.75$\pm$0.38 & 96.25$\pm$0.58 & 90.90$\pm$0.57 \\
4 & 99.84$\pm$0.31 & 98.18$\pm$0.52 & 93.59$\pm$0.83 & 100.00$\pm$0.00 & 98.91$\pm$0.53 & 96.95$\pm$1.05 & 98.44$\pm$0.35 & 95.47$\pm$1.12 & 89.02$\pm$1.30 & 99.06$\pm$0.53 & 96.30$\pm$0.91 & 91.02$\pm$0.69 \\
5 & 100.00$\pm$0.00 & 98.65$\pm$0.26 & 94.53$\pm$0.62 & 100.00$\pm$0.00 & 99.17$\pm$0.30 & 97.58$\pm$0.63 & 98.59$\pm$0.31 & 95.89$\pm$0.78 & 89.45$\pm$1.20 & 99.06$\pm$0.31 & 96.51$\pm$0.94 & 91.52$\pm$1.27 \\
6 & 99.92$\pm$0.16 & 98.59$\pm$0.42 & 95.04$\pm$0.73 & 100.00$\pm$0.00 & 99.11$\pm$0.21 & 97.54$\pm$0.47 & 98.52$\pm$0.57 & 96.15$\pm$0.45 & 89.65$\pm$1.37 & 98.83$\pm$1.02 & 96.46$\pm$0.27 & 91.52$\pm$0.56 \\
7 & 100.00$\pm$0.00 & 98.80$\pm$0.54 & 95.08$\pm$1.00 & 100.00$\pm$0.00 & 99.06$\pm$0.27 & 97.85$\pm$0.98 & 98.67$\pm$0.53 & 95.63$\pm$0.98 & 90.43$\pm$0.95 & 98.91$\pm$0.16 & 96.61$\pm$1.09 & 91.37$\pm$1.61 \\
8 & 100.00$\pm$0.00 & 98.96$\pm$0.37 & 95.27$\pm$1.26 & 100.00$\pm$0.00 & 99.01$\pm$0.38 & 97.58$\pm$0.75 & 98.75$\pm$0.29 & 95.89$\pm$1.05 & 90.00$\pm$1.62 & 98.98$\pm$0.53 & 96.67$\pm$0.92 & 91.88$\pm$0.81 \\
\hline
\end{tabular}
\label{tab:results_512}
\end{table}

\begin{table}[h]
\centering
\caption{Results on MNIST and Fashion-MNIST Datasets (Test set: 1024, Training set: 256)}
\footnotesize
\setlength{\tabcolsep}{4pt}
\begin{tabular}{c|cc|cc}
\hline
\multirow{2}{*}{Qubits} & \multicolumn{2}{c|}{MNIST} & \multicolumn{2}{c}{Fashion-MNIST} \\
\cline{2-5}
& 2-class & 3-class & 2-class & 3-class \\
\hline
3 & 99.80±0.12 & 97.63±0.38 & 97.85±0.21 & 95.23±0.34 \\
4 & 99.96±0.08 & 98.26±0.44 & 98.16±0.26 & 95.29±0.66 \\
5 & 100.00±0.00 & 98.78±0.27 & 98.28±0.15 & 95.63±0.78 \\
6 & 100.00±0.00 & 98.88±0.34 & 98.28±0.36 & 95.70±0.70 \\
7 & 100.00±0.00 & 98.83±0.22 & 98.32±0.23 & 95.60±0.63 \\
8 & 100.00±0.00 & 98.80±0.21 & 98.36±0.10 & 95.55±0.66 \\
\hline
\end{tabular}
\label{tab:results_1024}
\end{table}

\begin{table}[h]
\centering
\caption{Performance Comparison of Magnitude-Based and Our Quantum Attention Methods on MNIST and Fashion-MNIST.}
\begin{tabular}{|c|cc|cc|}
\hline
\multirow{2}{*}{Qubits} & \multicolumn{2}{c|}{MNIST (\%)} & \multicolumn{2}{c|}{Fashion-MNIST (\%)} \\
\cline{2-5}
& Overlap-Based \cite{zhao2024qksan,chen2025quantum} & Ours & Overlap-Based \cite{zhao2024qksan,chen2025quantum} & Ours \\
\hline
3 & 97.14 $\pm$ 0.70 & 97.24 $\pm$ 0.35 & 95.31 $\pm$ 0.89 & 95.63 $\pm$ 0.97 \\
4 & 97.97 $\pm$ 0.73 & 98.18 $\pm$ 0.52 & 95.36 $\pm$ 1.29 & 95.47 $\pm$ 1.12 \\
5 & 98.49 $\pm$ 0.30 & 98.65 $\pm$ 0.26 & 95.52 $\pm$ 0.98 & 95.89 $\pm$ 0.78 \\
6 & 98.49 $\pm$ 0.65 & 98.59 $\pm$ 0.42 & 95.78 $\pm$ 0.92 & 96.15 $\pm$ 0.45 \\
7 & 98.75 $\pm$ 0.56 & 98.80 $\pm$ 0.54 & 95.57 $\pm$ 0.79 & 95.63 $\pm$ 0.98 \\
8 & 98.80 $\pm$ 0.39 & 98.96 $\pm$ 0.37 & 95.63 $\pm$ 0.42 & 95.89 $\pm$ 1.05 \\
\hline
\end{tabular}
\label{tab:ablation}
\end{table}

\section{Circuit Architecture}
\label{section: Circuit Architecture}
\begin{figure}[H]
  \centering
  \includegraphics[width=3.3in]{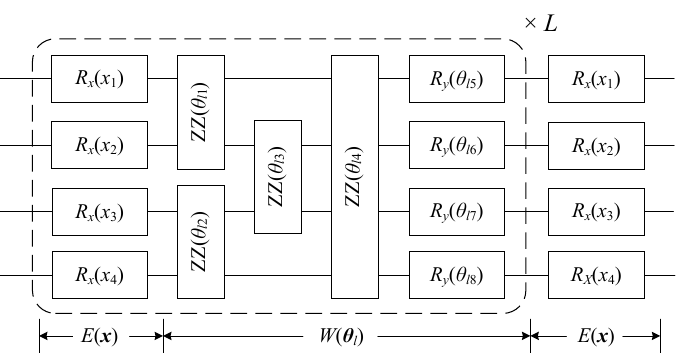}
  \caption{The architecture of a 4-qubit quantum feature mapping module.}
  \label{Figure Data Emembedding}
  \end{figure}
  
Fig. \ref{Figure Data Emembedding} illustrates the structure of the proposed quantum state embedding circuit architecture. The circuit uses an initial single-qubit $R_x$ gate for data encoding, followed by layers of parameterized $R_y$ gates and $ZZ$ gates to progressively enhance entanglement between qubits. These encoding and training structures can be extended by stacking L layers. The final $R_x$ gate completes the data mapping. 

\begin{figure}[H]
  \centering
  \includegraphics[width=3.3in]{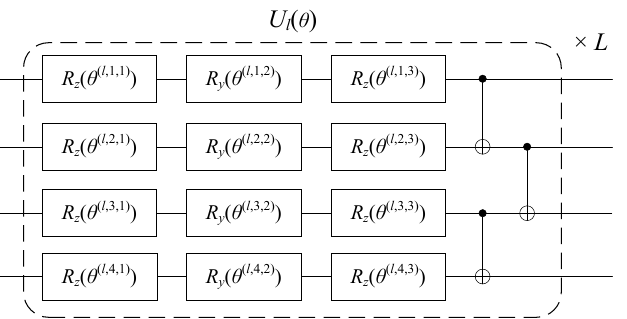}
  \caption{The architecture of a 4-qubit QFFN.}
  \label{Figure QFFN}
  \end{figure}
  
Fig. \ref{Figure QFFN} illustrates the hardware-efficient quantum circuit architectureused in the quantum feedforward neural network. Each layer consists of a sequence of $R_z$ and $R_y$ rotation gates applied to each qubit, followed by an entanglement layer formed by CNOT gates. This structure can be repeated $L$ times to enhance the complexity and expressiveness of the quantum states.

\vfill

\end{document}